Article type: Full Paper

# Highly-Efficient Selective Metamaterial Absorber for High-Temperature Solar Thermal Energy Harvesting

*Hao Wang,[1] Vijay Prasad Sivan,[2] Arnan Mitchell,[2]*
*Gary Rosengarten,[3] Patrick Phelan,[1] and Liping Wang[1,*]*

[1] School for Engineering of Matter, Transport & Energy,
Arizona State University, Tempe, AZ, 85287, USA

[2] School of Electrical and Computer Engineering, Royal Melbourne
Institute of Technology, Melbourne, VIC, 3001, Australia

[3] School of Aerospace, Mechanical and Manufacturing Engineering, Royal
Melbourne Institute of Technology, Carlton, VIC, 3053 Australia

*Corresponding author: Liping.Wang@asu.edu

**ABSTRACT**: In this work, a metamaterial selective solar absorber made of nanostructured titanium gratings deposited on an ultrathin $MgF_2$ spacer and a tungsten ground film is proposed and experimentally demonstrated. Normal absorptance of the fabricated solar absorber is characterized to be higher than 90% in the UV, visible and, near infrared (IR) regime, while the mid-IR emittance is around 20%. The high broadband absorption in the solar spectrum is realized by the excitation of surface plasmon and magnetic polariton resonances, while the low mid-IR emittance is due to the highly reflective nature of the metallic components. Further directional and polarized reflectance measurements show wide-angle and polarization-insensitive high absorption within solar spectrum. Temperature-dependent spectroscopic characterization indicates that the optical properties barely change at elevated temperatures up to 350 °C. The solar-to-heat conversion efficiency with the fabricated metamaterial solar absorber is predicted to be 78% at 100 °C without optical concentration or 80% at 400 °C with 25 suns, and could be further improved with better fabrication processes and geometric optimization during metamaterial design. The strong spectral selectivity, favorable diffuse-like behavior, and excellent thermal stability make the metamaterial selective absorber promising for significantly enhancing solar thermal energy harvesting in various system at mid to high temperatures.

## 1. Introduction

Clean and abundant solar energy has been intensively explored as an alternative to traditional fossil fuels over the past decades.[1] As a key component, absorbers that convert solar radiation into thermal energy greatly affect the performance of various solar thermal systems. An ideal solar absorber should possess an absorptance of unity in the solar spectrum covering UV, visible and near infrared (NIR) to convert most solar radiation into heat, along with zero emittance in the mid-IR regime to minimize energy loss from spontaneous thermal radiation.[2] This spectral selectivity is vital for solar thermal absorbers to achieve high solar-to-heat conversion efficiency. In addition, angular and polarization independence is highly desired for efficient solar absorbers considering the random nature of solar radiation. Excellent thermal stability is also crucial for ensuring solar absorbers to operate properly with efficient solar energy harvesting at elevated temperatures over time. Commercially, $TinO_x$[3] and Pyromark[4] have been used as solar absorbers for low- to high-temperature applications. Spectrally-selective TinOx coatings could absorb 95% of incident solar radiation and emit only 4% thermal radiation, but its performance is optimal only around 100°C. On the other hand, Pyromark exhibits a near-normal absorptance above 95% at high temperatures around 650°C, but its thermal emittance is also as high as 80%. Unfortunately, efficient solar absorbers with both spectral selectivity and high-temperature compatibility are still lacking.

Optical metamaterials refer to artificial structures with exotic optical properties that cannot be obtained in naturally occurring materials.[5] Selective absorption has been investigated in metamaterials made of different micro/nanostructures from GHz to IR spectral regime, including split-ring-resonators,[6, 7] fishnet,[8] cut-wires[9, 10] and photonic crystals.[11] Recently, film-coupled metamaterials in metal-insulator-metal configurations have been intensively studied as thermal emitters or selective absorbers from visible to NIR spectral regime. Wang and Zhang proposed a diffuse-like spectrally-selective thermophotovoltaic (TPV) emitter with 1D tungsten gratings on top of a SiO$_2$ spacer and tungsten substrate,[12] while Zhao et al. studied the polarization-independent 2D counterparts.[13] Note that both selective TPV emitters can be also used as selective solar absorbers due to the spectral selectivity well-matched to solar spectrum. Photonic crystals[14, 15] and nanoparticles[16-18] were also proposed for selective absorption in visible and NIR range. Highly-efficient solar absorbers require a broad absorption band from visible to NIR range, which could be attained with different geometric approaches. Wang and Wang proposed selective metamaterial solar absorbers with multi-sized tungsten patch arrays.[19] Aydin et al. discussed an ultra-thin broadband selective absorber composed of trapezoid gratings.[20] In addition, Lee et al. numerically explored a film-coupled nickel concave grating as a selective absorber for solar thermal energy harvesting.[21]

Most of previous experimental work on metamaterial absorbers has focused on optical characteristics at room temperature. Chen et al. demonstrated an IR metamaterial selective absorber with film-coupled cross-bar structure at room temperature.[22] Hao et al. investigated a plasmonic NIR selective absorber but neglected the effects of temperature on material properties.[23] However, the temperature-dependent optical properties of novel metamaterial solar absorbers have to be characterized in order to clearly understand the temperature effect as well as thermal stability, which is crucial for ensuring efficient solar energy harvesting at elevated temperatures.[24] Yeng et al. measured the spectral radiance at high temperature for a selective emitter made of tungsten photonic crystal,[25] but temperature dependence of radiative properties like absorptance or emittance is not investigated experimentally. Liu et al. explored the emittance of a metamaterial thermal emitter at varied temperatures but only in infrared range.[26] Recently,

MIT researchers demonstrated a thermally stable selective solar absorber made of 2D metallic dielectric photonic crystal structures after 24 hrs heating at 1000°C, but the temperature-dependent optical properties at high temperatures were not measured.[27] So far, the optical and radiative properties of selective solar absorbers at elevated temperatures have been little experimentally studied. The state of the art of the micro/nanostructured selective solar absorber was recently reviewed by Khodasevych et al. [28]

In this work, we report on the spectroscopic characterization at both room and elevated temperatures of a selective metamaterial solar absorber made of a 2D titanium grating deposited on an $MgF_2$ spacer and an opaque tungsten film, as illustrated in **Figure 1**a. Tungsten is chosen as the substrate material due to its excellent high-temperature stability, while titanium is selected for the gratings as it is easier to pattern with lift-off process than tungsten. The near-normal specular and hemispherical reflectance is measured over a broad spectral range from UV to the mid-IR regime, demonstrating its spectral selectivity. The electromagnetic field distribution obtained from finite-difference time-domain (FDTD) simulation is plotted at absorption peaks to explain underlying mechanism. The effects of oblique incidence, polarization state, as well as temperature up to 350ºC are further experimentally investigated. Finally, the solar-to-heat conversion efficiency for the metamaterial solar absorber is predicted to show its excellent performance especially at higher absorber temperatures.

## 2. Results and Discussion

### 2.1. Optical and Radiative Properties at Near Normal Incidence

The proposed metamaterial solar absorber is fabricated using electron-beam lithography, thin-film deposition, and lift-off processes (see Experimental Section). Figure 1b shows the photo of the tested metamaterial solar absorber sample with a 5.4 mm by 5.4 mm pattern area on a 21 mm by 18 mm Si wafer. The fabricated grating patterns at the top layer of the metamaterial solar absorber have excellent symmetry in x and y direction as seen from the top-view SEM image in Figure 1c, while a trapezoid shape is observed from the side-view SEM image in Figure 1d, which is typical for metallic gratings patterned from a lift-off process with negative photoresist. The measured geometric parameters are: grating period $\Lambda = 600$ nm, grating top width $w_1 = 200$ nm, bottom width $w_2 = 360$ nm, grating height $h = 170$ nm, and $MgF_2$ spacer thickness $t = 50$ nm. The tungsten layer has a thickness of 200 nm, which is opaque within the spectral region of interests.

The specular reflectance $R'_\lambda$ for the fabricated metamaterial solar absorber is measured using a Fourier Transform Infrared (FTIR) spectrometer, while the diffuse reflectance and hemispherical reflectance $R'^\cap_\lambda$ is characterized using an integrating sphere (IS) along with a tunable light source covering a broad spectral range from UV to NIR (i.e., 0.35 μm to 1.6 μm in wavelength). **Figure 2**a shows the room-temperature specular, diffuse, and hemispherical reflectance measured at different wavelengths with an incidence angle $\theta = 8°$ from both the FTIR and IS measurements. It can be observed that the specular reflectance $R'_\lambda$ of the metamaterial solar absorber is lower than 5% at 0.35 μm < $\lambda$ < 0.8 μm, while $R'_\lambda$ increases to a maximum value of 11% in the NIR range. The low reflectance indicates high absorptance in the solar spectrum, which is desired for highly-efficient solar absorbers.

The results of Figure 2a also show that the diffuse reflectance is negligible (less than 1%) at wavelengths $\lambda > 650$ nm, indicating that the sample is highly specular. This can be explained by the sub-wavelength nature of the metamaterial solar absorber array due to its periodicity. In periodic structures, larger in-plane wavevectors can be obtained with high-order diffracted waves by:

$$k_{\parallel,mn} = \left(k_{x,\text{inc}} + \frac{2\pi m}{\Lambda_x}\right)\hat{x} + \left(k_{y,\text{inc}} + \frac{2\pi n}{\Lambda_y}\right)\hat{y} \tag{1}$$

where $m$ and $n$ are the diffraction orders respectively in the x and y directions, while $\Lambda_x = \Lambda_y = 600$ nm is the grating period in both directions for the fabricated metamaterial solar absorber. At normal incidence (i.e., $k_{x,\text{inc}} = k_{y,\text{inc}} = 0$), when the wavelength of incident waves is larger than the grating period, $k_{\parallel,mn}$ is larger than the free-space wavelength $k_{\text{inc}} = 2\pi/\lambda$ for all the diffracted waves except for the zeroth order (i.e., $m = n = 0$), which is the specular reflection component. In other words, all the non-zero diffraction orders are evanescent waves in the sub-wavelength region, which do not contribute to the far-field diffuse reflection. As a result, the hemispherical reflectance $R_\lambda'^\cap$ from the IS measurement is almost the same as the specular reflectance $R_\lambda'$ from the FTIR measurement at longer wavelengths beyond 650 nm with a small difference less than 2.5% mainly due to the different signal-to-noise ratios in both measurements. This highlights the specular behavior of the metamaterial solar absorber when the incident wavelength is larger than the grating period at normal incidence.

On the other hand, when the incident wavelength is smaller than the grating period, the diffuse reflectance becomes significant and increases up to 6.5% around $\lambda = 0.5$ μm due to the non-negligible contribution from higher-order diffracted waves. As a result, the difference between hemispherical and specular reflectance becomes larger at short wavelengths due to the increased diffuse reflection. Note that from the IS measurement, the diffuse reflectance starts to become negligible at 650 nm, which is not the same as the grating period $\Lambda = 600$ nm. This is because the reflectance is measured at near-normal with $\theta = 8°$ (i.e.,) instead of perfectly normal incidence. The small nonzero oblique incidence angle results in $k_{x,\text{inc}} \neq 0$, thereby slightly shifting the cutoff wavelength to $\lambda = 650$ nm.

The spectral-directional absorptance of the metamaterial solar absorber can be obtained by $\alpha'_\lambda = 1 - R'^\cap_\lambda$ based on the energy balance as the sample is opaque, while the spectral-directional emittance is simply equal to the spectral-directional absorptance according to Kirchhoff's law: $\alpha'_\lambda = \varepsilon'_\lambda$. As discussed previously, the metamaterial solar absorber is highly specular at wavelengths $\lambda > 650$ nm. Thus, the spectral-hemispherical reflectance $R'^\cap_\lambda$ at $\lambda > 650$ nm is obtained as the specular reflectance (i.e., $R'^\cap_\lambda = R'_\lambda$) from the FTIR measurement at near-normal incidence, while $R'^\cap_\lambda$ at shorter wavelengths is acquired from the IS measurement. Figure 2b shows the characterized spectral absorptance/emittance of the fabricated metamaterial solar absorber under near-normal incidence at room temperature. The metamaterial solar absorber shows absorptance higher than 90% within 0.35 μm < $\lambda$ < 2 μm from the UV to NIR region, and emittance around 20% from 6 μm < $\lambda$ < 20 μm in the mid-IR. Therefore, the spectral selectivity of the metamaterial solar absorber is clearly demonstrated, which is crucial for improving the performance of solar absorbers by maximizing solar absorption and minimizing self-emission loss.

FDTD simulation was performed to numerically calculate the spectral-normal absorptance/emittance, which shows excellent agreement with the measurement data in Figure 2b. There are two major absorption peaks from the measurement for the metamaterial solar absorber: one at $\lambda = 0.68$ μm with an amplitude of 97.8% due to surface plasmon polariton (SPP),[2, 29, 30] and the other at $\lambda = 1.6$ μm with an amplitude of 94.1% due to magnetic polariton (MP).[12, 13, 19, 31] SPP is a surface wave due to the collective oscillation of plasmon excited at the interface of two materials with permittivity of $\varepsilon_1$ and $\varepsilon_2$. Note that the real parts of $\varepsilon_1$ and $\varepsilon_2$ should have opposite signs, and a SPP is excited when the following dispersion relation is satisfied:[2]

$$k_{SPP} = \frac{\omega}{c_0}\sqrt{\frac{\varepsilon_1 \varepsilon_2}{\varepsilon_1 + \varepsilon_2}} \qquad (2)$$

The dispersion relation theoretically predicts a SPP mode excited for the Ti gratings at $\lambda_{SPP} = 0.68$ μm with $\theta = 8°$. It can be observed that the measured absorptance is slightly higher than simulation results at the visible range of 0.4 μm $< \lambda <$ 0.6 μm. This is because the SPP absorption peak wavelength is highly dependent on incidence angles. During the measurement, the light is incident onto the sample at $\theta = 8°$ with a half-cone angle of 1.5° or so. Therefore, the absorption peak associated with SPP is broadened from the measurement in this visible regime due to the finite solid angle of the incident beam.

On the other hand, MP is the coupling between the incident electromagnetic wave and magnetic resonance inside the structure. To explain the excitation mechanism of MP, the electromagnetic field distribution at the MP resonance is obtained from the FDTD simulation and plotted in **Figure 3**. The arrows show the electric field vectors and the contour represents the strength of the magnetic field normalized as $\log_{10}|H/H_0|^2$, where $H_0$ is the incident magnetic field. It can be observed that the electric current forms a loop under the Ti patch, while the magnetic field is greatly enhanced in the local area within the current loop with one order of magnitude higher over the incidence. This is a typical electromagnetic field pattern at MP resonance due to the diamagnetic response inside the grating microstructure,[12, 13, 19, 31] and the strong field confinement explains the high absorption at MP resonance.

The measured emittance in the mid infrared is around 20%, which is higher than the 4% predicted from the simulation, possibly due to the oxidation of tungsten and titanium during the sample fabrication process. Note that, the metamaterial solar absorber is essentially highly reflective from the metallic components without any resonance absorption in the long-wavelength region. By addressing the material oxidation issue, it is expected that, the emittance of the metamaterial solar absorber sample can be further reduced to approach the theoretical 4% to better minimize the thermal emission energy loss, thereby further improving the solar thermal conversion performance.

**2.2. Optical and Radiative Properties at Oblique Incidences**

The optical and radiative properties at oblique angles of a solar absorber are also vital for efficiently harvesting direct sunlight coming from different directions after an optical concentrator. An ideal solar absorber should be diffuse-like with optical and radiative properties independent of direction. Therefore, the specular reflectance of the metamaterial solar absorber was measured by the FTIR at incidence angles $\theta$ = 5°, 15°, 30° and 45° at both TM and TE polarizations.

**Figure 4**a plots the measured specular reflectance $R'_\lambda$ for the metamaterial solar absorber at oblique TM incidence. It can be observed that $R'_\lambda$ at $\lambda < 0.7$ µm is lower than 5% for all oblique incidences. Moreover, the reflection dip due to the MP excitation at $\lambda$ = 1.6 µm does not shift with increased incidence angle, thanks to the unique direction-independent characteristic of MP resonance.[12, 31] When the incidence angle increases for TM incidence, the strength of the incident H field parallel to the y-direction grating groove does not change. Thus,

the strength of the oscillating current loop for MP does not decrease, and the MP resonance strength remains almost unchanged.[13] As a result, the reflectance at the MP wavelength increases little with larger incidence angle.

Figure 4b shows the specular reflectance $R'_\lambda$ of the metamaterial solar absorber at TE oblique incidence. It is found that $R'_\lambda$ at $\lambda < 0.7$ μm is also lower than 5%. The reflectance at MP resonance increases slightly with larger oblique angles. This is because the strength of incident H field component parallel to the grating groove in y direction decreases as the incidence angle becomes larger for TE waves. As a result, the strength for MP resonance decreases and the absorptance drops.[13] The reflectance at longer wavelengths in the mid-IR also increases slightly at oblique incidences.

Considering the random nature of sunlight, the reflectance for unpolarized waves, which is averaged from both polarizations, is presented in Figure 4c for different oblique angles. It is observed that when incidence angle $\theta$ changes from 5° to 30°, the reflectance for unpolarized incidence barely changes with reflectance lower than 15% in the visible and NIR region (i.e., 0.4 μm to 2 μm in wavelength). The reflectance slightly increases but remains less than 20% when $\theta$ further increases to 45°. The measured optical and radiative properties of the fabricated metamaterial solar absorber sample clearly demonstrate the diffuse-like behaviors at both TM and TE polarizations as well as unpolarized waves.

## 2.3. Optical Properties at Elevated Temperatures

In order to characterize the optical properties of the metamaterial solar absorber at elevated temperatures, an FTIR fiber optics technique as illustrated in **Figure 5**a was developed for measuring the temperature-dependent specular reflectance $R'_\lambda$ from the sample mounted inside a home-designed heater assembly with precise temperature control. Note that a 30-nm-thick SiO$_2$ layer was deposited onto the sample surface to protect the metamaterial structures from possible oxidation or chemical reaction in air during sample heating, which would change the desired optical properties and degrade the performance of the metamaterial solar absorber at elevated temperatures.

Figure 5b shows the reflectance for the metamaterial solar absorber when the sample temperature increases from room temperature (23.5°C) to 350°C with an interval of 50°C. It can be seen that the reflectance at wavelengths from 0.4 µm to 0.8 µm barely changes with increased absorber temperature, indicating excellent thermal stability of the fabricated metamaterial solar absorber. The reflectance from 0.8 µm < $\lambda$ < 2 µm decreases with higher temperatures, which might be due to materials chemical changes. Nevertheless, the slight variation of the reflectance is only within 2.5%, and decreased NIR reflectance at higher temperature is actually beneficial for absorbing more solar energy.

**2.4. Solar Thermal Efficiency Analysis with the Metamaterial Solar Absorber**

In order to evaluate the performance of the metamaterial as a potential highly-efficient solar thermal absorber, the solar-to-heat conversion efficiency is theoretically analyzed. Assuming no conduction or convection losses, the conversion efficiency of a solar absorber can be calculated by:

$$\eta = \frac{\alpha_{\text{Total,N}} CG - \varepsilon_{\text{Total,N}}(\sigma T_{\text{A}}^4 - \sigma T_{\text{sky}}^4)}{CG} \tag{3}$$

where $C$ is the concentration factor, $G$ is the heat flux of incident solar irradiation at AM1.5 (global tilt),[32] $T_A$ is the absorber temperature, and $T_{sky} = 20°C$ is the sky temperature. $\alpha_{Total,N}$ and $\varepsilon_{Total,N}$ are respectively the total normal absorptance and emittance for the solar absorber, which can be respectively calculated by:

$$\alpha_{Total,N} = \int_0^\infty \alpha'_{\lambda,N} I_{AM1.5}(\lambda) d\lambda \bigg/ \int_0^\infty I_{AM1.5}(\lambda) d\lambda \tag{4}$$

$$\varepsilon_{Total,N} = \int_0^\infty \varepsilon'_{\lambda,N} I_{BB}(\lambda, T_A) d\lambda \bigg/ \int_0^\infty I_{BB}(\lambda, T_A) d\lambda \tag{5}$$

where $I_{AM1.5}(\lambda)$ is the spectral intensity of solar irradiation at AM1.5 (global tilt),[32] $I_{BB}(\lambda, T_A)$ is the spectral blackbody radiative intensity at the solar absorber temperature $T_A$, and $\alpha'_{\lambda,N}$ and $\varepsilon'_{\lambda,N}$ are respectively the spectral normal absorptance and emittance of the solar absorber measured at room temperature. Note that both $\alpha'_{\lambda,N}$ and $\varepsilon'_{\lambda,N}$ are taken to be independent on temperature as observed from the temperature-dependent optical characterization. For the calculation of total absorptance, the spectral integration is limited to the wavelength region from 0.35 μm to 4 μm, because our instrument cannot measure optical properties at wavelengths below 0.35 μm while the available AM1.5 data only covers wavelengths up to 4 μm. There is still around 7% of solar radiation outside of this spectral range mainly in the UV regime. Similarly, the spectral integration for total emittance is performed in the wavelengths from 0.35 μm to 20 μm limited by the available measurement data. Note that, there is only 4% energy outside this spectral regime mainly in the far-infrared for a blackbody with a temperature of 400°C. Since the metamaterial solar absorber is quite diffuse with oblique angles $\theta < 45°$ from the directional optical property characterization, the total hemispherical absorptance or emittance can be reasonably approximated by the total normal absorptance or emittance.

**Figure 6**a plots the conversion efficiency $\eta$ as a function of absorber temperature $T_A$ under 1 sun (i.e., no optical concentration) for an ideal selective surface, the metamaterial solar absorber with optical and radiative properties taken from either measurements or the FDTD simulation, and a black surface. The absorptance for the ideal surface is unity below the cutoff wavelength to maximize absorbed solar radiation, while its emittance is zero beyond the cutoff wavelength to minimize spontaneous thermal emission loss. The cutoff wavelength for the ideal selective solar absorber is optimized at each absorber temperature for maximal conversion efficiency, which represents the upper limit. On the other hand, the black surface has unity absorptance and emittance in the entire spectral regime (i.e., $\varepsilon = \alpha = 1$), whose conversion efficiency indicates the lowest limit.

It is observed that the conversion efficiency for the metamaterial solar absorber with measured optical properties could reach 78.1% at the absorber temperature $T_A = 100°C$ and monotonically drops to zero at the stagnation temperature of 241°C, at which no solar thermal energy is harvested. The efficiency for the metamaterial absorber with simulated optical properties shows relatively higher values. Theoretically, the proposed metamaterial absorber could have a conversion efficiency as high as 88.3% at $T_A = 100°C$ and a much higher stagnation temperature of 393°C. The discrepancy on the efficiency results from larger emittance in the mid-IR region from the measurement than simulation. The performance of the fabricated metamaterial absorber can be further improved to approach the theoretical values after the oxidation issues during the sample fabrication are addressed.

In comparison, a black surface could only convert about 32% of solar energy to useful heat at $T_A = 100°C$, while its efficiency drops quickly to zero at 125°C, suggesting the great importance of spectral selectivity in enhancing the solar-to-heat conversion efficiency. On the other hand, the efficiency of the metamaterial absorber is about 10% (with simulated optical properties) or 20% (with measured data) less than the ideal surface at $T_A = 100°C$, mainly due to the larger emittance in the mid-IR around 4% (simulated) and 20% (measured). The absorptance within solar spectrum of the metamaterial is also smaller than the ideal by 5% to 10%, but at low optical concentrations, the self-emission loss determined by the mid-IR emittance plays a major role in determining the amount of harvested solar energy. The difference between the metamaterial absorber and the ideal surface becomes even larger when temperature goes up as the ideal surface maintains high efficiencies of 98.7% at 200°C, and 93% at 400°C. This is because the cut-off wavelength of the ideal surface is optimized at each temperature, as the blackbody spectrum governed by the Planck's law would shift towards shorter wavelengths with higher absorber temperature. On the other hand, the cut-off wavelength for the metamaterial absorber is around $\lambda = 2$ μm or so and does not change with absorber temperature. In fact, the cut-off wavelength of the metamaterial absorber is determined by the MP resonance wavelength, which can be easily tuned with geometric parameters such as grating width. Therefore, for a given absorber temperature required by a particular solar thermal system, the cut-off wavelength as well as the absorptance/emittance spectrum can be optimized during the design and fabrication processes for achieving maximal solar-to-heat conversion efficiency.

It is known that with concentrated sunlight, the solar-to-heat conversion efficiency can be further improved. Here, we consider the effect of concentration factor $C$ from 1 to 50 at an absorber temperature $T_A = 400°C$ for a medium-temperature application. Note that the thermal energy at 400°C carries quite amount of exergy, and could potentially deliver electricity with heat engines in Rankin cycle or solid-state devices such as thermoelectrics and TPV besides heating and cooling applications. Figure 6b shows that, the metamaterial absorber with measured optical properties could harvest 21.5 % of solar energy to useful heat under 5 suns, 57.4 % with 10 suns, and 80% with 25 suns. For the metamaterial absorber with even lower IR emittance as simulated, the efficiency could be as high as 71% under 5 suns and 81% with 10 suns, indicating room for improvement with the current fabricated sample. With more optical concentrations up to 100 suns, the conversion efficiency of the metamaterial absorber (with optical properties both measured and simulated) saturates towards 90%. In comparison, the efficiency of the ideal surface slightly increases from 93% to 98% at the same temperature from 1 sun to 100 suns, while that of a black surface is improved greatly with more optical concentrations from 0% at 11 suns toward 90% with 100 suns, suggesting that the spectral selectivity becomes less important with a factor larger than 100 as thermal emission loss becomes negligible with highly concentrated incident solar radiative flux.

## 3. Conclusion

In summary, we have proposed and experimentally characterized the optical and radiative properties of a highly-efficient metamaterial selective absorber for solar thermal energy harvesting at elevated temperatures up to 350°C. The metamaterial solar absorber is demonstrated to exhibit absorptance higher than 90% within the solar spectrum and low emittance of around 20% in the mid IR regime. Besides, the measured reflectance shows diffuse-like behavior at large oblique incidence, indicating wide-angle and polarization-insensitive high absorption within solar spectrum. Characterization of the temperature-dependent optical properties of the solar absorber sample demonstrated excellent thermal stability and temperature-insensitivity up to 350°C. Based on the measured optical properties of the fabricated metamaterial sample, the solar-to-heat conversion efficiency from this solar absorber is predicted to be 78% at 100°C without optical concentration, or 80% at 400°C with 25 suns. By addressing the oxidation issues during fabrication, the efficiency from the metamaterial solar absorber could be increased by a few more percent, indicated by the simulated data. In addition, by carefully optimizing the geometry parameters from the design, the efficiency could be further maximized for a given operating temperature. Insights gained from this work will facilitate the design and optimization of a new class of selective solar absorbers for enhancing solar energy harvesting at medium-to-high temperatures in various systems including but not limited to solar heating and cooling, concentrated solar power, solar thermoelectrics, and solar TPV.

## 4. Experimental Section

### 4.1. Sample Fabrication

The metamaterial selective solar absorber is fabricated with the following procedure. First, $MgF_2$ and tungsten thin films were deposited using e-beam evaporation (Kurt J. Lesker PVD 75) on a silicon substrate. Then, the 2D Ti gratings with period of 600 nm were fabricated onto the $MgF_2$/W coated Si substrate using electron beam lithography by a multi-step exposure scheme on a FEI Nova Nano SEM with NPGS (J. C. Nabity Lithography Systems, Nanometer Pattern Generation System), followed by e-beam evaporation and lift-off process. Large-area patterns of 5.4 mm × 5.4 mm were finally obtained through above fabrication procedure.

### 4.2. Specular Reflectance Measurement with FTIR

Specular reflectance for the solar absorber was measured by the FTIR (Thermo Fisher iS50) along with a variable-angle reflectance accessory (Harrick Scientific, Seagull). For near normal measurement, the specular reflectance was measured at an incidence angle of $8°$ from 0.4 to 20 µm in wavelength with a spectral resolution of 4 cm$^{-1}$ in wavenumber. The radiative properties at TM and TE incidences are almost identical at $8°$ due to the excellent geometric symmetry of the fabricated metamaterial solar absorber. Therefore, the measurement was performed for unpolarized waves. For measurement at oblique incidence, the specular reflectance was measured at incidence angles of 5°, 15°, 30° and 45°. The reflectance is measured separately for TM and TE waves with either a broadband polarizer (Thorlabs, WP25M-UB) in visible and NIR regime or the internal wire-grid IR polarizer inside the FTIR. The reflectance from 0.4 to 1.1 µm in wavelength was measured by an internal Si detector, while an internal DTGS detector was employed at longer wavelengths beyond 1.1 µm. An aluminum mirror was used as the reference and the measured reflectance is normalized based on the theoretical reflectance of aluminum. The measured reflectance was averaged over three measurements (each with 32 scans) by interchanging the sample and reference to reduce the occasional errors during the measurement. In order to obtain the uncertainty of FTIR measurements, the reflectance of reference Si sample (Virginia Semiconductor, Boron doped (110) Si sample, with resistivity of 60 ohms-cm) was measured and compared with its theoretical value, confirming the measurement uncertainty to be within 2%.

**4.3. Directional-Hemispherical and Diffuse Reflectance Characterization with IS**

The directional-hemispherical and diffuse reflectance was measured in the integrating sphere (Labsphere) at an incidence angle of $8°$ from 0.35 µm to 1.6 µm. An unpolarized monochromatic light from UV to NIR was provided by the tunable light source (Newport, TLS-250Q) with a spectral resolution of 10 nm, and the broadband reflectance was obtained by taking a wavelength scan and measuring the reflectance at each wavelength with a Si detector (Thorlabs, SM05PD1A) and an InGaAs detector (Thorlabs, SM05PD5A). A silver mirror was employed as the reference and the measured reflectance is corrected with the theoretical reflectance of silver. The reflectance was averaged from five individual measurements. The measurement uncertainty for IS was also proved to be within 2% with reference Si sample.

**4.4. Optical Characterization at Elevated Temperature with FTIR Fiber Optics**

The experimental setup for the FTIR fiber optics measurement is shown in Figure 5a. An FTIR fiber coupler (Harrick, Fibermate2) was employed to couple the FTIR with a VIS-NIR reflection (or called back-scattering) optics fiber bundle (Thorlabs, RP21). A fiber probe with collimating and focusing optics yielded a beam spot with a diameter of 4 mm onto the sample surface. The normally reflected signal was then collected by the same probe and acquired by the FTIR detectors through optical fibers. The sample was mounted onto a copper disk inside a home-built heater assembly. A thermocouple (Omega, KMTXL-040) was utilized to measure the sample temperature and the temperature was sent to a temperature controller (Omega, CSi8D), which modulated the power input to a custom-designed heater assembly and thereby accurately maintained the sample temperature at the setpoint within 1 K variation. The FTIR measurements were performed when the desired temperature was stable at least for 20 min. Note that in the NIR spectral regime, the measured reflectance at each wavelength is averaged from 20 neighboring data points to reduce the fluctuation in measured reflectance caused by the low signal-to-noise ratio from the DTGS detector. The measurement uncertainty was found to be smaller than 2.5% with reference Si.

**4.5. FDTD simulation**

The FDTD simulation was performed with a commercial package (Lumerical Solutions, FDTD solutions). Optical properties of titanium, MgF$_2$ and tungsten were obtained from Palik's data.[33] The simulation was implemented in a 0.6 μm × 0.6 μm × 4 μm simulation domain, and the wavelength range of interest is from 0.3 μm to 20 μm with a spectral resolution of 5 nm. Manual refined mesh with size of 5 nm in x and y directions, and 2 nm in z direction was employed to ensure the convergence for numerical simulation. Periodic boundary conditions were applied in x and y directions for normal incidence, while perfect matched layers with reflection coefficient of 10$^{-6}$ were placed in z direction. A plane wave source was set 1.2 μm above the structure surface, and the reflectance $R$ was obtained by a frequency-domain power monitor positioned 0.5 μm above the plane wave source. The spectral absorptance was obtained using $\alpha = 1 - R_\lambda'^\cap$ as the structure is opaque due to the 200 nm tungsten substrate.

**ACKNOWLEDGMENT**

This work was supported by the *US-Australia Solar Energy Collaboration - Micro Urban Solar Integrated Concentrators (MUSIC)* project sponsored by the *Australian Renewable Energy Agency (ARENA)*. HW and LW also would like to thank the support from the ASU New Faculty Startup fund. The authors acknowledge the facilities, and the scientific and technical assistance of the Australian Microscopy & Microanalysis Research Facility at RMIT University.

**Figure Captions:**

**Figure 1.** (a) Structure schematic for proposed metamaterial solar absorber. $\mathbf{K}_{inc}$ represents the incident wavevector, and the incidence angle $\theta$ is defined as the angle between $\mathbf{K}_{inc}$ and the surface normal. The plane of incidence (POI) is defined as the x-z plane. The $\mathbf{E}$ field is in the POI for transverse magnetic (TM) incidence, while it is perpendicular to the POI for transverse electric (TE) incidence. (b) A photo of the fabricated sample for optical characterization. SEM images of the fabricated absorber sample from (c) top view and (d) side view.

**Figure 2.** (a) Measured room-temperature specular, diffuse, and hemispherical reflectance of the metamaterial solar absorber. (b) Measured and simulated room-temperature spectral absorptance of the metamaterial solar absorber.

**Figure 3.** Cross-sectional view of electromagnetic field at the MP resonance wavelength from the FDTD simulation. The arrows represent strength and direction of the electric field, while the contour plot illustrates strength of magnetic field.

**Figure 4.** (a) Specular reflectance of the metamaterial solar absorber sample measured by FTIR at (a) oblique TM incidences, (b) oblique TE incidences, and (c) unpolarized oblique incidences.

**Figure 5.** (a) Schematic of the experimental setup for the temperature-dependent FTIR fiber optics measurements for characterizing spectral normal reflectance at elevated temperatures. (b) Specular reflectance of the metamaterial solar absorber measured at elevated temperatures up to 350°C with temperature-dependent FTIR fiber optics.

**Figure 6.** (a) Predicted solar-to-heat conversion efficiencies of an ideal selective surface, the metamaterial solar absorber (with optical properties either measured or simulated), and a black surface as a function of absorber temperature $T_A$ under 1 sun. (b) Solar-to-heat conversion efficiency for all three surfaces as a function of concentration factor $C$ at an absorber temperature of $T_A = 400°C$.

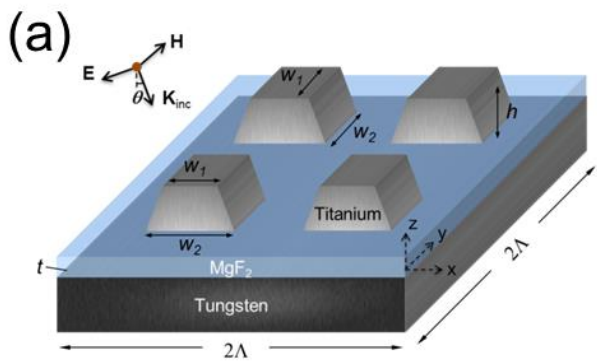 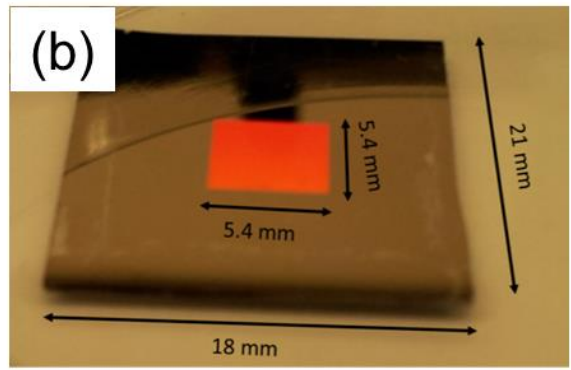
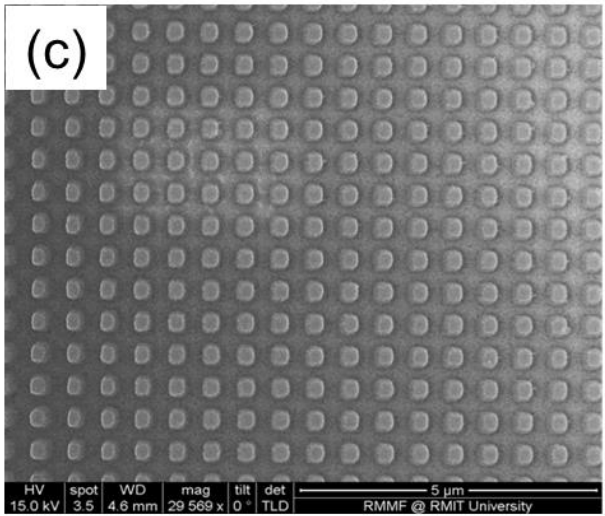 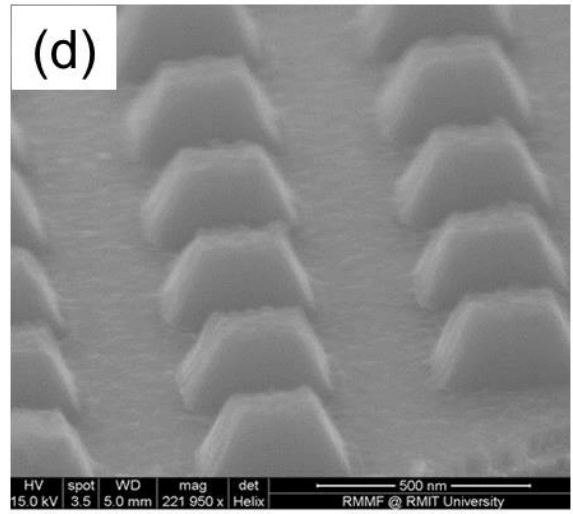

Wang, et al., Figure 1

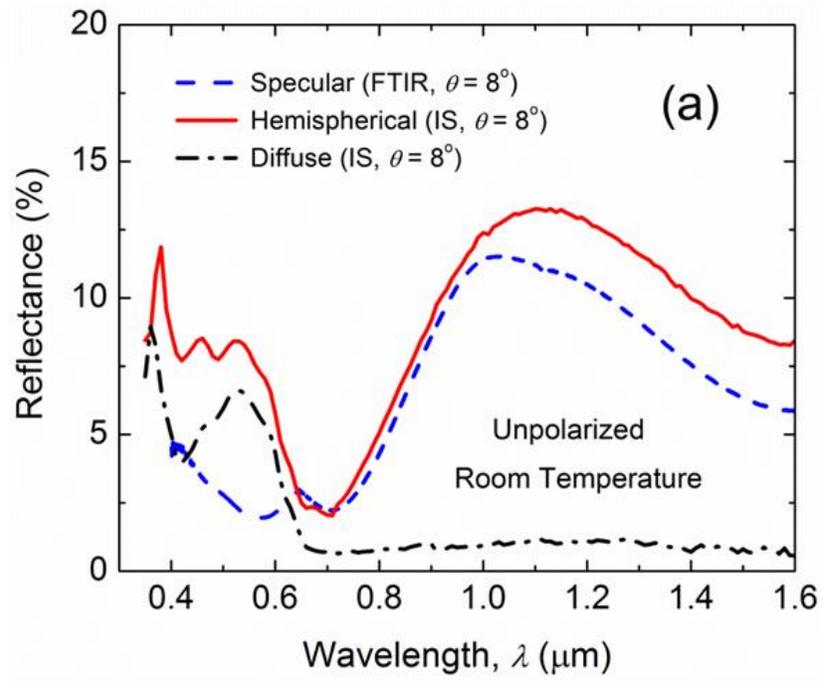

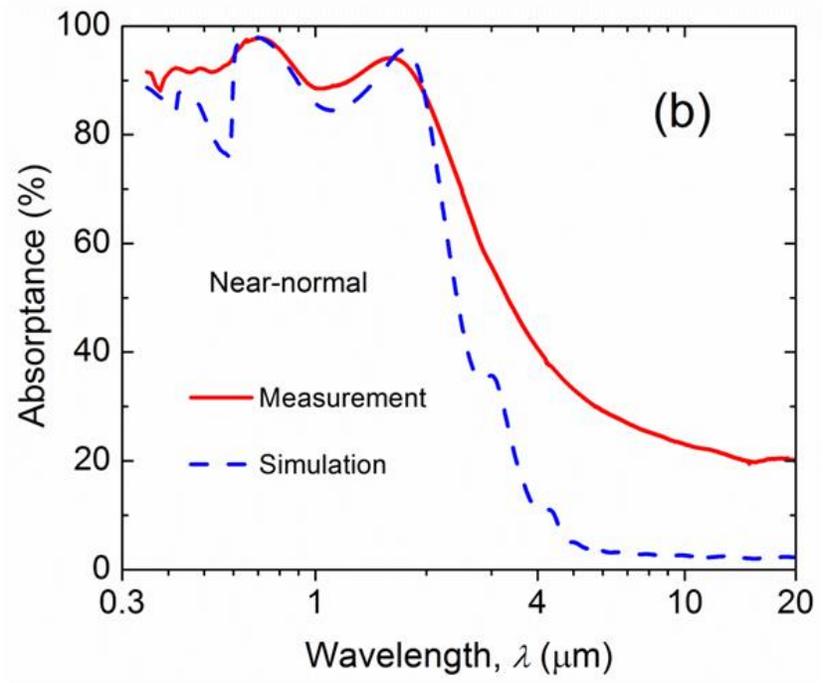

Wang, et al., Figure 2

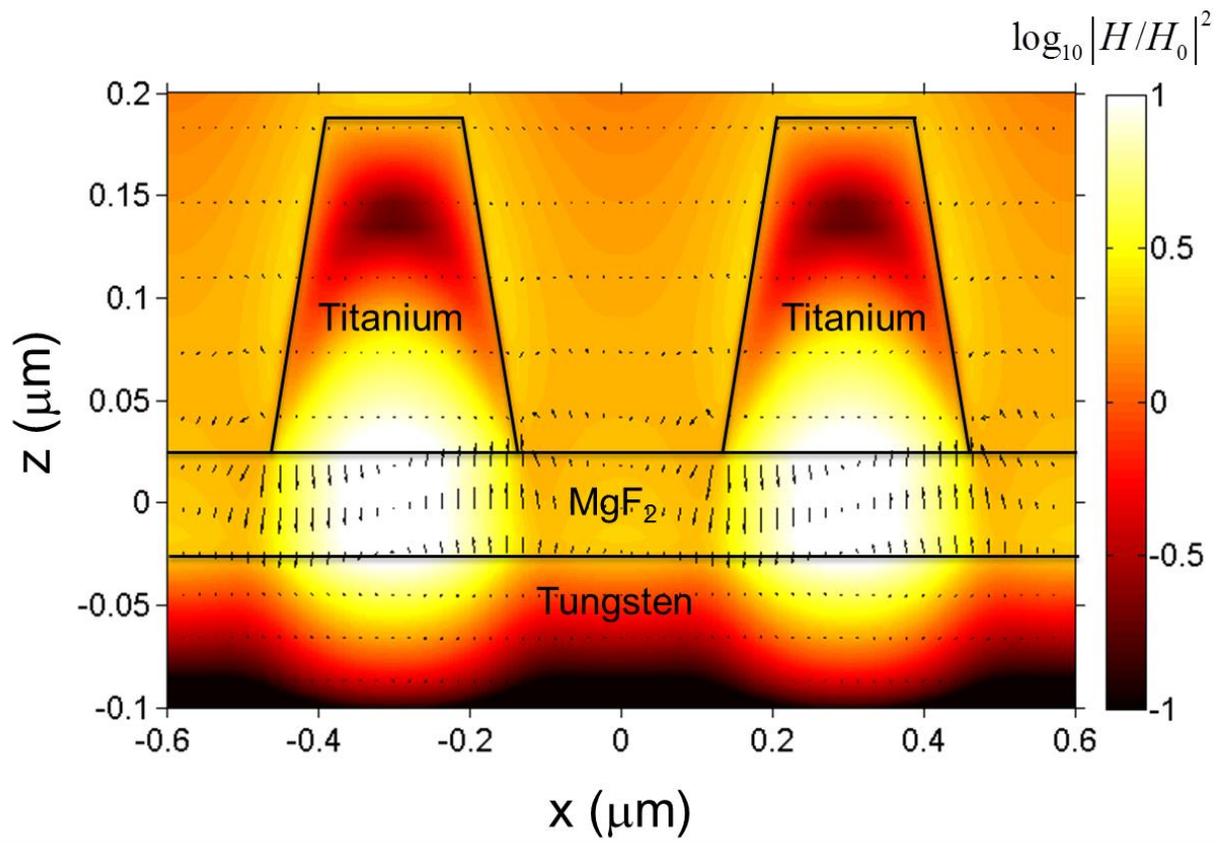

Wang, et al., Figure 3

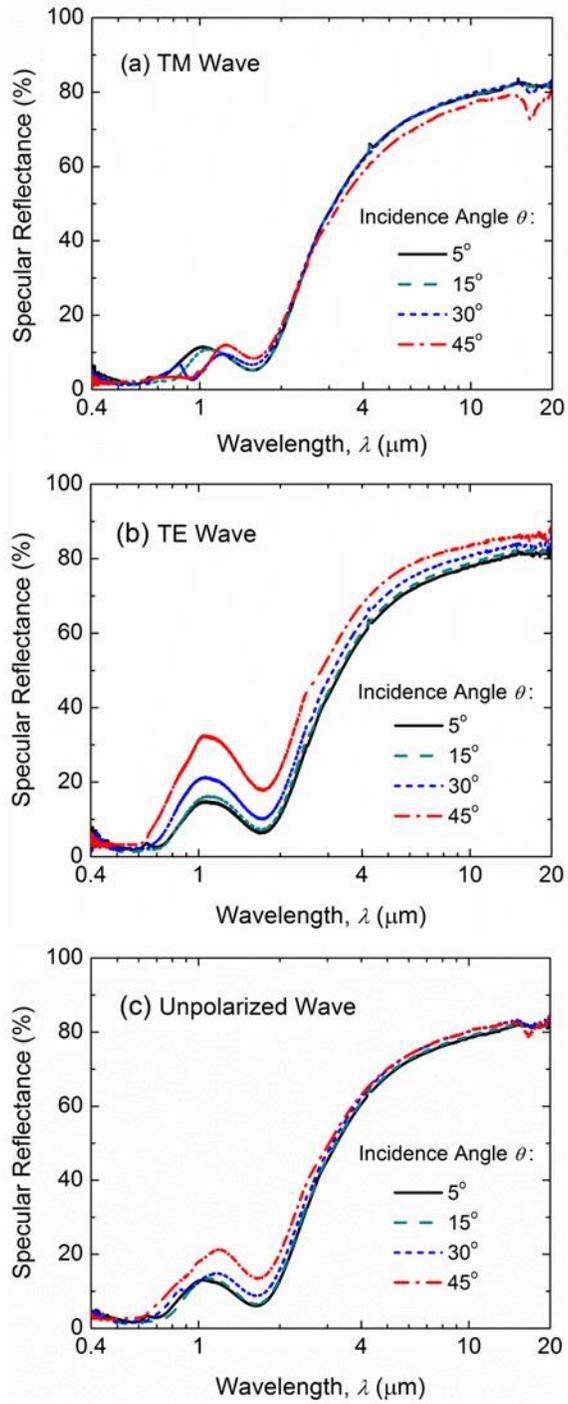

Wang, et al., Figure 4

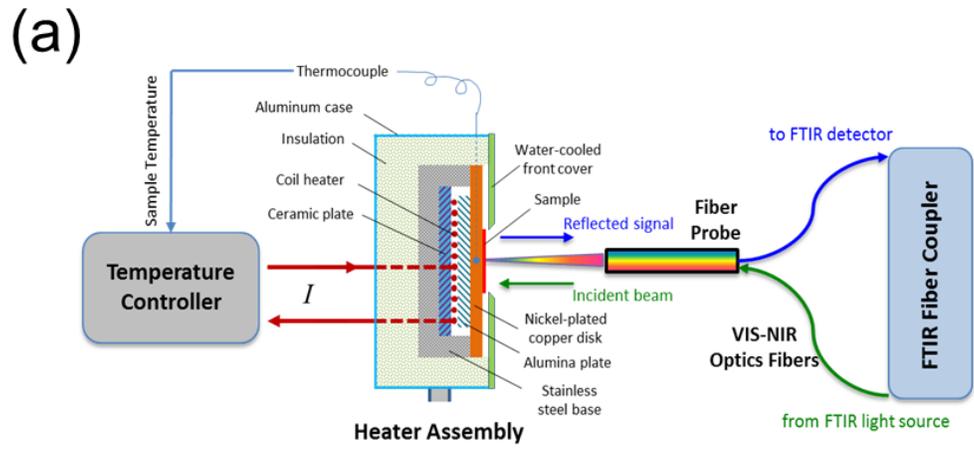

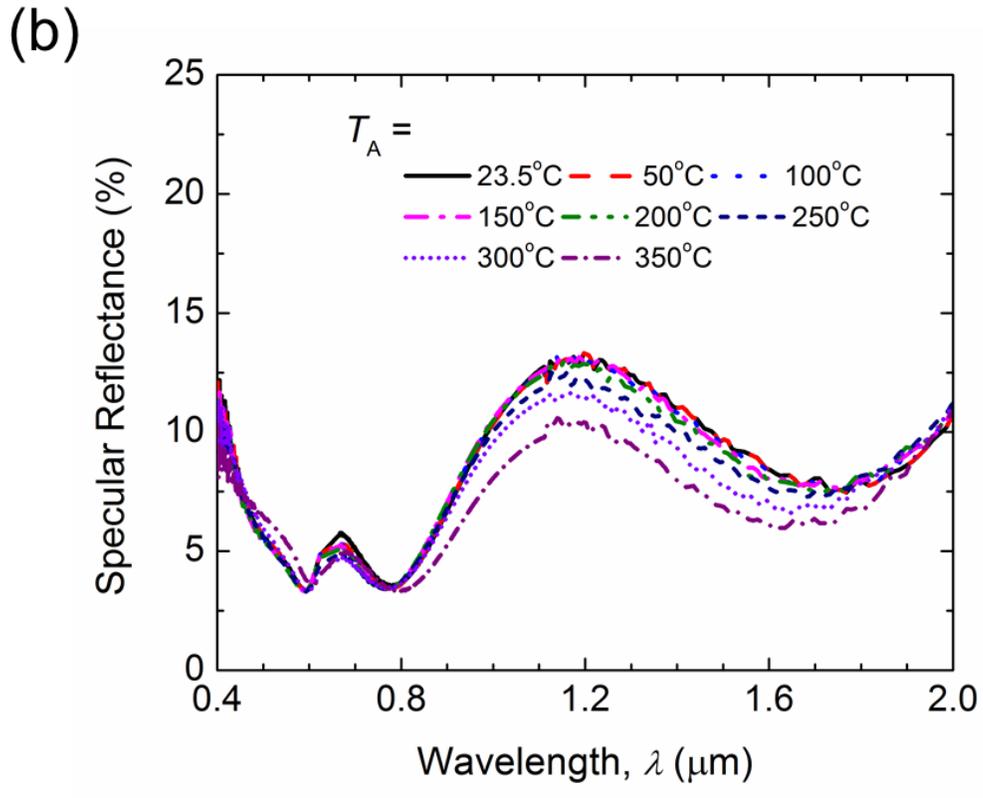

Wang, et al., Figure 5

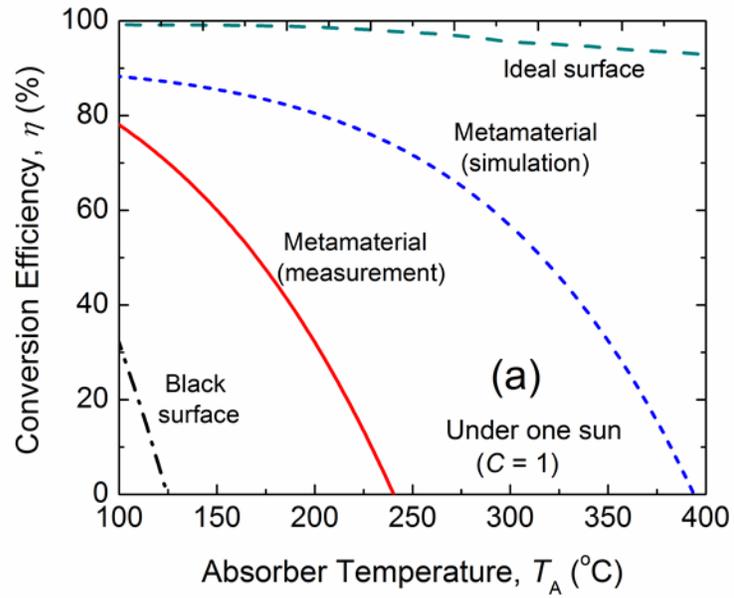

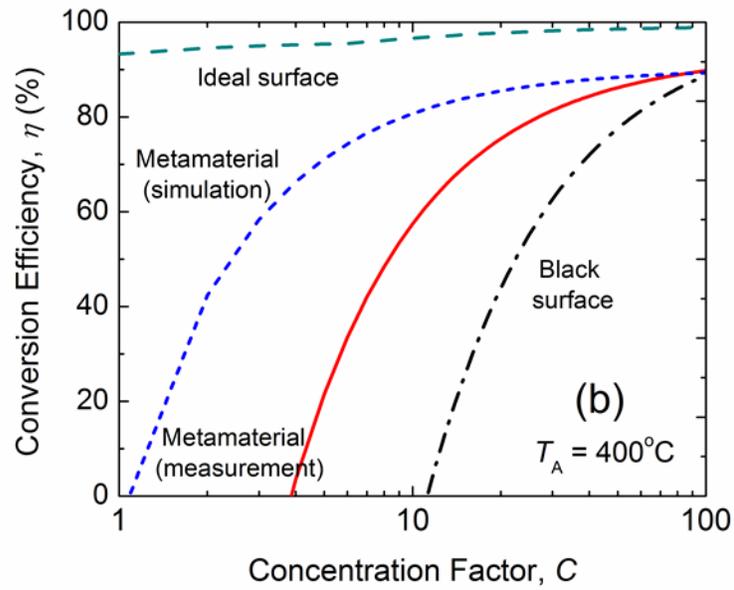

Wang, et al., Figure 6